# Improved Material Decomposition with a Two-step Regularization for spectral CT

Weiwen Wu[1], Peijun Chen[1], Vince Vardhanabhuti[2], Weifei Wu[1] and Hengyong Yu[3], Senior Member, IEEE

[1]The First People's Hospital of Yichang, Yichang, 443000, China
[2] Department of Diagnostic Radiology, University of Hong Kong, Hong Kong, 999077, China
[3] Department of Electrical and Computer Engineering, University of Massachusetts Lowell, Lowell, MA 01854, USA

The contribution of W.W. Wu and P.J. Chen are equal

Corresponding authors: W.F Wu (e-mail: wuweifei236@sina.com) and H.Y Yu (hengyong-yu@ieee.org).

This work was supported in part by the National Natural Science Foundation of China (No. 81702198).

**ABSTRACT** One of the advantages of spectral computed tomography (CT) is it can achieve accurate material components using the material decomposition methods. The image-based material decomposition is a common method to obtain specific material components, and it can be divided into two steps: image reconstruction and post material decomposition. To obtain accurate material maps, the image reconstruction method mainly focuses on improving image quality by incorporating regularization priors. Very recently, the regularization priors are introduced into the post material decomposition procedure in the iterative image-based methods. Since the regularization priors can be incorporated into image reconstruction and post image-domain material decomposition, the performance of regularization by combining these two cases is still an open problem. To realize this goal, the material accuracy from those steps are first analyzed and compared. Then, to further improve the accuracy of decomposition materials, a two-step regularization based method is developed by incorporating priors into image reconstruction and post material decomposition. Both numerical simulation and preclinical mouse experiments are performed to demonstrate the advantages of the two-step regularization based method in improving material accuracy.

## I. INTRODUCTION

The traditional computed tomography (CT) may share the same or very similar CT numbers for the materials that have different elemental compositions. This becomes the biggest challenge for differentiating different types of tissues [1]. As a solution, spectral CT was developed. A spectral CT system can employ a photon counting detector (PCD) to collect projections at different energy bins. The PCD can discriminate the carried energy with incident photon by setting several energy thresholds. A spectral CT system can obtain multiple-energy projections, which is beneficial to differentiate material components and determine tissue distributions within the imaged object [2]. For example, it can be applied to differentiate small (≤30mm) hepatic hemangioma (HH) from small hepatocellular carcinoma (HCC) [3], characterize kidney stone [4], image contrast media [5], improve tumor visibility in the vicinity of gold fiducial markers [6], determine urinary stone composition [7] and left ventricular thrombus diagnosis [8]. In fact, projections obtained by a spectral CT system have a lower signal-to-noise ratio (SNR) than that obtained by the conventional CT in theory. Due to the effects of x-ray fluorescence, charge sharing, K-escape and pulse pileups, the projections are tarnished by unknown noise [9] and the accuracy of material decomposition is compromised.

To obtain higher accuracy for the decomposed materials, it is necessary to develop advanced material decomposition methods by considering physical effects and more regularized priors. Generally speaking, the material decomposition methods can fall into two classes: direct and indirect methods [10]. The direct material decomposition algorithms can achieve material maps directly [11, 12] by exploring an accurate x-ray emitting spectrum. However, because the accuracy of estimated x-ray spectrum can be influenced by the scatter effect, the response of detector and so on [13], it is difficult to accurately model and estimate the x-ray spectrum in practice [14, 15]. Although the regularization-based material reconstruction model [16] is a feasible strategy to improve the accuracy of decomposed materials to some extent, the material decomposition results are sensitive to noise.

Regarding the indirect material decomposition methods, they can further be divided into the projection-based and image-based methods [17]. In the projection-based methods, the raw measurements can be decomposed into specific material measurements and then final material images can be reconstructed by certain reconstruction techniques [18, 19]. However, the measured noise can be magnified during the course of measurement decomposing, which can compromise the material accuracy. Regarding the image-based techniques, the implementation can be divided into two steps: image reconstruction and post material decomposition. Higher quality of reconstructed images is beneficial to the improvement of material accuracy. A natural idea for improving material accuracy is to achieve higher quality of reconstructed images. To achieve this goal, a lot of iterative image reconstruction algorithms were developed by incorporating regularized priors. For example, total variation (TV) minimization is a common constraint to enhance the sparsity of spectral CT images from each energy channel [20]. To exploit data redundancies in the energy domain, the Local HighlY constrained backPRojection Reconstruction (HYPR-LR) was proposed [21]. The tight frame sparsity was employed to spectral breast CT [22]. Motivated by the PCA model [23], the prior rank, intensity and sparsity model (PRISM) was studied [24]. From the view of energy spectrum, spectral CT imaging is very similar to hyperspectral images [25, 26], and images from different energy channels share similar image structures and features. Here , many other algorithms were developed, including the patch-based low-rank reconstruction [27], tensor dictionary learning (TDL)[28] and its modified version ($L_0$TDL)[29], spatial-spectral cube matching frame (SSCMF) [30], non-local low-rank cube tensor factorization (NLCTF) [31] and aided by self-similarity in image-spectral tensors (ASSIST) [32]. To further improve the accuracy of material images, it is natural to incorporate high quality prior into the reconstruction model, such as the spectral prior image constraint compressed sensing (SPICCS)[33], the average-image-incorporated block-matching and 3D (aiiBM3D) filtering [34], nonlocal spectral similarity model [35], prior image constrained total generalized variation [36], the total image constrained diffusion tensor (TICDT) [37].

The image reconstruction step can help to achieve high accuracy of decomposed materials by improving image reconstruction quality. This is because high quality of reconstructed images is good for post image-domain material decomposition [28]. However, the material decomposition accuracy may be limited. To further improve material accuracy, numerous iterative image-domain material decomposition (IID-MD) methods were proposed for dual-energy CT [38-41]. For example, Niu *et al.* first proposed an iterative full variance-covariance matrix of material components based material decomposition method by introducing the quadratic smoothness penalty function to constrain the feasibility of material images [42]. Then, non-local mean [40], spectral diffusion[43], nonlinear decomposition [44], similarity-based regularization (PWLS-SBR)[45], entropy minimization[46], data-driven sparsity [47], multiscale penalized weighted least-squares [48], fully convolutional network [49] and PWLS-TNV-$\ell_0$ [50] were further proposed. However, these methods are mainly developed for dual energy CT rather than spectral CT. Usually, there are only two basis materials within the imaging object for dual energy CT. When the number of material is three or more, the IID-MD model for dual-energy cannot be employed to spectral CT material decomposition. To address this issue, Tao *et al.* [51] developed a prior knowledge aware iterative denoising material decomposition (MD-PKAD) model to obtain basis material images. Xie *et al.* established a multiple constraint image-domain material decomposition (MCIMD) model and a numerical phantom was employed to validate its performance [52]. Very recently, the dictionary learning-based image-domain material decomposition (DLIMD) was proposed by our group [53].

The regularization priors in either image reconstruction model or post material decomposition model can improve the accuracy of material decomposition. However, the performance of improving material accuracy by combining these two cases is still an open problem. Therefore, it is significant to compare the benefits of the regularization prior in the reconstruction model or post material decomposition model in practice. Besides, the strategy of two-step regularization is also explored to further improve the accuracy of material decomposition based on the analysis of the advantages for regularization in different steps.

The rest of this paper is organized as follows. In section II, the spectral CT image reconstruction and image-domain material decomposition models for spectral CT will be presented. In section III, both numerical simulation and real dataset experiments are performed to make a quantitative and qualitative comparison and analysis. In section IV, we will discuss some related issues and make a conclusion.

## II. MATERIAL AND METHOD

### A. SPECTRAL CT IMAGE RECONSTRUCTION

Limited by x-ray radiation dose [54], scan range [55] , system noise [56] and so on, the x-ray spectral CT image reconstruction is a typical ill-posed inverse problem, and its results are unstable and irreversible [57]. Because the emitting x-ray spectrum is divided into several energy bins, the projections including a series of energy channels can be obtained with one scan. Without loss of generality, we consider the fan-beam geometry, and the x-ray CT image reconstruction can be characterized as a discrete linear equation set problem. We have

$$\boldsymbol{h}_n = \boldsymbol{A}\boldsymbol{p}_n + \boldsymbol{\epsilon}_n, n = 1, \dots, N, \quad (1)$$

where $\boldsymbol{A} \in \mathbb{R}^{L \times J}$ ( $L = L_1 \times L_2$ and $J = J_1 \times J_2$ ) represents the system matrix, $L_1$ and $L_2$ are the number of projection view and detector element, $J_1$ and $J_2$ represent the width and height of the reconstructed image, $n$ ($n = 1,2,\cdots,N$) represents the index for energy-channel, $N$ is the number of energy-channel, $\boldsymbol{h}_n \in \mathbb{R}^J$ ($n = 1,2,\cdots,N$) is a vectorized $n^{th}$ energy-channel image, $\boldsymbol{p}_n \in \mathbb{R}^L$ is the vectorized projection of $n^{th}$ energy bin and $\epsilon_n \in \mathbb{R}^L$ is the corresponding measurement noise. Since the system matrix $A$ is too huge to operate in practice, Eq. (1) can be iteratively solved by the following linear programming problem:

$$\underset{\mathcal{H}}{\text{argmin}} \sum_{n=1}^{N} \frac{1}{2} \|p_n - Ah_n\|_F^2, \quad (2)$$

where $\mathcal{H} \in \mathbb{R}^{J_1 \times J_2 \times N}$ is a 3$^{\text{rd}}$–order tensor and $h_n$ represents the $n^{th}$ row of the mode-3 unfolding of $\mathcal{H}$. To reconstruct high-quality spectral CT images, a common strategy is to incorporate the regularization prior. Considering the regularization prior, Eq. (2) can be rewritten as

$$\underset{\mathcal{H}}{\text{argmin}} \sum_{n=1}^{N} \frac{1}{2} \|p_n - Ah_n\|_F^2 + \frac{\eta}{2} \mathcal{L}(\mathcal{H}), \quad (3)$$

where $\eta > 0$ is a regularization parameter and $\mathcal{L}(\cdot)$ represents the general regularization form. To optimize the objective function Eq. (3), we first introduce $\mathcal{S}$ to replace $\mathcal{H}$ and then it can be converted into the following two sub-problems

$$\underset{\mathcal{H}}{\text{argmin}} \left( \sum_{n=1}^{N} \frac{1}{2} \|p_n - Ah_n\|_F^2 \right) + \frac{\tau}{2} \|\mathcal{H} - \mathcal{S}^{(k)}\|_F^2, \quad (4a)$$

$$\underset{\mathcal{S}}{\text{argmin}} \frac{1}{2} \|\mathcal{S} - \mathcal{H}^{(k+1)}\|_2^2 + \frac{\eta}{2\tau} \mathcal{L}(\mathcal{S}). \quad (4b)$$

Since Eq. (4a) is convex, it can be easily solved by using the steepest descent method. As for Eq. (4b), different regularizers have different optimization method. All the regularizers can be divided into two categories: channel-dependent and channel-independent. The channel-dependent methods can consider the correlation of different channel images $\{h_n\}_{n=1}^N$ in the reconstruction models, such as low-rank [28], non-local similarity [30], *etc*. In this case, Eq. (4b) should be optimized for all the channels simultaneously. For channel-independent methods (i.e., TV-based model [20], aiiBM3D[34], (SPICCS)[33]), the correlation of different channel images is ignored in these models, and Eq. (4b) is equal to

$$\underset{\{S_n\}_{n=1}^N}{\text{argmin}} \sum_{n=1}^{N} \frac{1}{2} \|S_n - H_n^{(k+1)}\|_F^2 + \sum_{n=1}^{N} \mu_n \mathcal{L}(S_n), \quad (5)$$

where $\mu_n$ is the regularization parameter for $n^{th}$ energy bin, $S_n$ and $H_n$ represent the matrix forms of $h_n$ and $s_n$.

**B. IMAGE-DOMAIN MATERIAL DECOMPOSITION**

The goal of material decomposition is to obtain material maps $\{m_v\}_{v=1}^V$ ($v$ represents the index of basis material and $V$ is the number of basis materials) from the reconstructed channel images $h_n (1 \leq n \leq N)$. The image-domain material decomposition model can be expressed as

$$\begin{bmatrix} b_{11} & \cdots & b_{V1} \\ \vdots & \ddots & \vdots \\ b_{1N} & \cdots & b_{VN} \end{bmatrix} \begin{bmatrix} (m_1)^T \\ \vdots \\ (m_V)^T \end{bmatrix} = \begin{bmatrix} (h_1)^T \\ \vdots \\ (h_N)^T \end{bmatrix}, \quad (6)$$

where $b_{vn}$ represents the mean attenuation coefficients from $v^{th}$ basis material of $n^{th}$ energy-channel and its computation method is given in [58], and $(.)^T$ represents the transpose operation. Eq. (6) can be simplified as

$$B\mathcal{M}_{(3)} = \mathcal{H}_{(3)}, \quad (7)$$

where $= \begin{bmatrix} b_{11} & \cdots & b_{VN} \\ \vdots & \ddots & \vdots \\ b_{T1} & \cdots & b_{VN} \end{bmatrix} \in \mathbb{R}^{V \times N}$ is material mean attenuation coefficients matrix, $m_v$ is the $v^{th}$ row of the mode-3 unfolding of $\mathcal{M}$. According to Eq. (7), the material images can be obtained by the direct inversion (DI) method [38]

$$\mathcal{M}_{(3)} = (B^T B)^{-1} B^T \mathcal{H}_{(3)}. \quad (8)$$

Unfortunately, the relationship between material attenuation coefficients and x-ray emitting spectrum is non-linear, and the process of material decomposition is unstable. In addition, the reconstructed spectral CT images usually contain noise, which can compromise the material decomposition accuracy. Considering noise, Eq. (7) can be rewritten as

$$B\mathcal{M}_{(3)} = \mathcal{H}_{(3)} + e, \quad (9)$$

where $e \in \mathbb{R}^{V \times N}$ represents the noise. Eq. (9) can be converted into the following problem

$$\underset{\mathcal{M}}{argmin} \frac{1}{2} \|B\mathcal{M}_{(3)} - \mathcal{H}_{(3)}\|_F^2, \#(10)$$

where $\|\cdot\|_F$ is the Frobenius norm. To further improve the material accuracy, the regularization priors can be introduced into Eq. (10) to constrain the solution. Then, we have

$$\underset{\mathcal{M}}{argmin} \frac{1}{2} \|B\mathcal{M}_{(3)} - \mathcal{H}_{(3)}\|_F^2 + \frac{\lambda}{2} \mathcal{L}_1(\mathcal{M}), \#(11)$$

where $\lambda > 0$ is a regularization factor to balance data fidelity term $\frac{1}{2} \|B\mathcal{M}_{(3)} - \mathcal{H}_{(3)}\|_F^2$ and regularization term $\mathcal{L}_1(\mathcal{M})$.

To further improve the final material decomposition accuracy, the volume conservation is introduced into Eq. (11) as constrain by considering the air as a basis material $\mathcal{N}$. Then we have,

$$\left( \sum_{v=1}^{V} \mathcal{M}_{j_1 j_2 v} \right) + \mathcal{N}_{j_1 j_2} = 1 \ (1 \leq j_1 \leq J_1, 1 \leq j_2 \leq J_2), (12)$$

where $\mathcal{M}_{j_1 j_2 v}$ and $\mathcal{N}_{j_1 j_2}$ represents the $(j_1, j_2, v)^{th}$ and $(j_1, j_2)^{th}$ pixel value of $\mathcal{M}$ and $\mathcal{N}$. Besides, all elements of $\mathcal{M}$ should be in the range of [0 1], *i.e.*,

$$0 \leq \mathcal{M} \leq 1. \quad (13)$$

By introducing the aforementioned two constraints, the general regularized image-domain material decomposition model [53] can be established as

$$\underset{\mathcal{M}}{min} \left( \frac{1}{2} \|B\mathcal{M}_{(3)} - \mathcal{H}_{(3)}\|_F^2 + \frac{\lambda}{2} \mathcal{L}_1(\mathcal{M}) \right),$$

$$s.t. \ 0 \leq \mathcal{M} \leq 1, \left( \sum_{v=1}^{V} \mathcal{M}_{j_1 j_2 v} \right) + \mathcal{N}_{j_1 j_2} = 1$$

$$(1 \leq j_1 \leq J_1, 1 \leq j_2 \leq J_2). \quad (14)$$

To optimize Eq. (14), we first introduce $\mathcal{W}$ to replace $\mathcal{M}$ and then it is converted to the following two sub-problems:

$$\min_{\mathcal{M}} \left( \frac{1}{2} \|\mathbf{B}\mathcal{M}_{(3)} - \mathcal{H}_{(3)}\|_F^2 + \frac{\delta}{2} \|\mathcal{W}^{(k)} - \mathcal{M}\|_F^2 \right),$$

$$s.t. \, \mathbf{0} \leq \mathcal{M} \leq \mathbf{1}, \left( \sum_{v=1}^{V} \mathcal{M}_{j_1 j_2 v} \right) + \mathcal{N}_{j_1 j_2} = 1$$

$$(1 \leq j_1 \leq J_1, 1 \leq j_2 \leq J_2), \quad (15)$$

$$\min_{\mathcal{W}} \frac{\delta}{2} \|\mathcal{W} - \mathcal{M}^{(k+1)}\|_F^2 + \frac{\lambda}{2} \mathcal{L}_1(\mathcal{W}), \quad (16)$$

where δ stands for coupling factor between data fidelity term and error feedback term. Regarding Eq. (15), it is equal to the following problem

$$\min_{\mathcal{M}} \left( \frac{1}{2} \|(\mathbf{B}^T \mathbf{B} + \delta \mathbf{I})\mathcal{M}_{(3)} - (\mathbf{B}^T \mathcal{H}_{(3)} + \delta(\mathcal{W}^{(k)})_{(3)})\|_F^2 \right),$$

$$s.t. \, \mathbf{0} \leq \mathcal{M} \leq \mathbf{1}, \left( \sum_{v=1}^{V} \mathcal{M}_{j_1 j_2 v} \right) + \mathcal{N}_{j_1 j_2} = 1$$

$$(1 \leq j_1 \leq J_1, 1 \leq j_2 \leq J_2). \quad (17)$$

Eq. (17) can be solved using a similar strategy in [53]. As what we discussed in [53], there are two strategies to realize this goal. If using strategy 1 to optimize Eq. (17), the air can increase the number of basis materials, and it can result in instability of material decomposition and degrade the accuracy of material decomposition. The results in [53] have demonstrated this conclusion in noise-free case. Thus, the strategy 2 is used in this study. For Eq. (16), if the correlation of different materials is ignored, each material can be updated independently

$$\min_{\{\mathbf{W}_n\}_{n=1}^{N}} \sum_{n=1}^{N} \left( \frac{1}{2} \|\mathbf{W}_n - \mathbf{M}_n^{(k+1)}\|_F^2 + \frac{\lambda_n}{2} \mathcal{L}_1(\mathbf{W}_n) \right). \quad (18)$$

It can be divided into the following $N$ sub-problems

$$\min_{\mathbf{W}_n} \left( \frac{1}{2} \|\mathbf{W}_n - \mathbf{M}_n^{(k+1)}\|_F^2 + \frac{\lambda_n}{2} \mathcal{L}_1(\mathbf{W}_n) \right), n = 1, \ldots, N, \quad (19)$$

where $\lambda_n$ represents the regularization parameter of $n^{\text{th}}$ material. The optimization of Eq. (19) depends on the specific form of $\mathcal{L}_1(\cdot)$.

### C. ALGORITHM IMPLEMENTATION

To compare material decomposition accuracy of the same regularization coupling with spectral CT image reconstruction model and image-domain post material decomposition model, a representative regularized prior should be incorporated. Because the image objects in our numerical simulations and preclinical experiment are approximately piecewise constants, the total variation (TV) [59] minimization is selected as the regularization for implementation [51]. Here, the material decomposition from four methods, *i.e,* simultaneous algebraic reconstruction technique (SART) followed by DI (SART-DI), TV Minimization (TVM) followed by DI (TVM-DI), SART followed by TV-based material decomposition TVMD (SART-TVMD) and TVM followed by TVMD (TVM-TVMD) are compared. More details about TVM and TVMD can refer to [31] and [53]. All parameters in all methods are optimized in this study. To make a quantitative analysis for material decomposition accuracy, three metrics (root means square error (RMSE), structural similarity (SSIM) and peak-signal-to-noise ratio (PSNR)) are employed.

## III. EXPERIMENTS AND RESULTS

### A. NUMERICAL SIMULATIONS

An injected 1.2% iodine contrast agent numerical mouse thorax phantom is utilized to implement simulation study. **Fig. 1** shows the mouse thorax phantom, which contains three materials (*i.e.,* bone, soft tissue and iodine contrast agent). A polychromatic 50 kVp x-ray source is divided into 4 energy bins ([16, 22) keV, [22, 25) keV, [25, 28) keV, [28, 50] keV), as shown in **Fig. 2**. A PCD with 512 cells (each element covers 0.1*mm*) is used to collect 640×4 projections for a full scan. The distance between X-ray source and PCD is 180 *mm* and that between X-ray source and object is 132 *mm*. The photon number of each emitting x-ray path is set as $5 \times 10^3$. Here, Poisson noise is superimposed to projections directly. The reconstructed images by the SART and TVM methods consist of 512×512×4 pixels, and they are given in **Fig. 3**. Here, the number iteration is set as 30 for all image reconstruction and material decomposition. Compared with the SART, the TVM can improve image quality using regularized prior and suppress the noise.

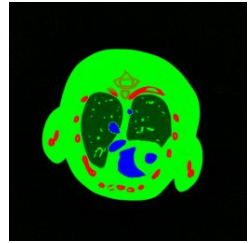

**Figure 1. Mouse thorax phantom, where green, red and blue stand for water, bone and iodine contrast agent, respectively.**

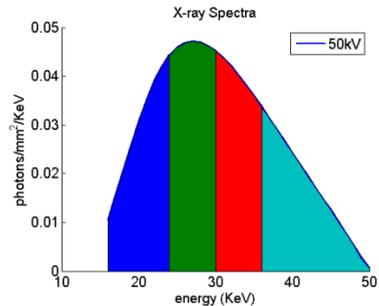

**Figure 2. The used 50 keV x-ray emitting spectrum.**

To evaluate the material decomposition accuracy of all methods in the case of numerical mouse, **Fig. 4** shows the results of three basis materials (bone, soft tissue and iodine contrast agent) using SART-DI, TVM-DI, SART-TVMD and TVM-TVMD. Here, the ground truth of three basis materials is obtained from the SART results using DI in case of noise-free projections. From **Fig. 4**, one can see many iodine contrast agent pixels are wrongly classified as bone and the accuracy of bone is decreased a lot in the SART-DI results. Because the reconstructed images using

SART contain noise, the final material decomposition results are compromised. Compared with the SART-DI results, the TVM-DI and SART-TVMD methods can provide better quality of material images. More pixels of iodine contrast agent are wrongly classified into bone in the SART-TVMD results than those of the TVM-DI. However, the accuracy of material component can be further improved by the TVM-TVMD method, which is confirmed by the magnified ROI A in **Fig.4**. In terms of soft tissue, the lung structures and features from the SART-DI are submerged by noise. Although the TVM-DI can obtain high quality of material image, the blocky artifacts appear around the image features and noise is still obvious. Since the TV constraint penalizes material image directly, the results by the SART-TVMD are smoother than those obtained by the TVM-DI. Because the TVM-TVMD inherits the advantages of both SART-TVMD and TVM-DI, it can improve material accuracy in image reconstruction and material decomposition and further gain high quality soft tissue images. These conclusions can be confirmed by the magnified ROIs B, C and D in **Fig. 5**. As for the iodine contrast agent, not only the results of SART-DI contain noise but also many pixels of bone are classified into iodine. Therefore, the accuracy is severe compromised. Compared with the SART-DI, TVM-DI and SART-TVMD can improve material accuracy that less pixels are wrongly treated as iodine contrast agent. Compared with the TVM-DI and SART-TVMD methods, the TVM-TVMD can further improve the accuracy of iodine contrast agent.

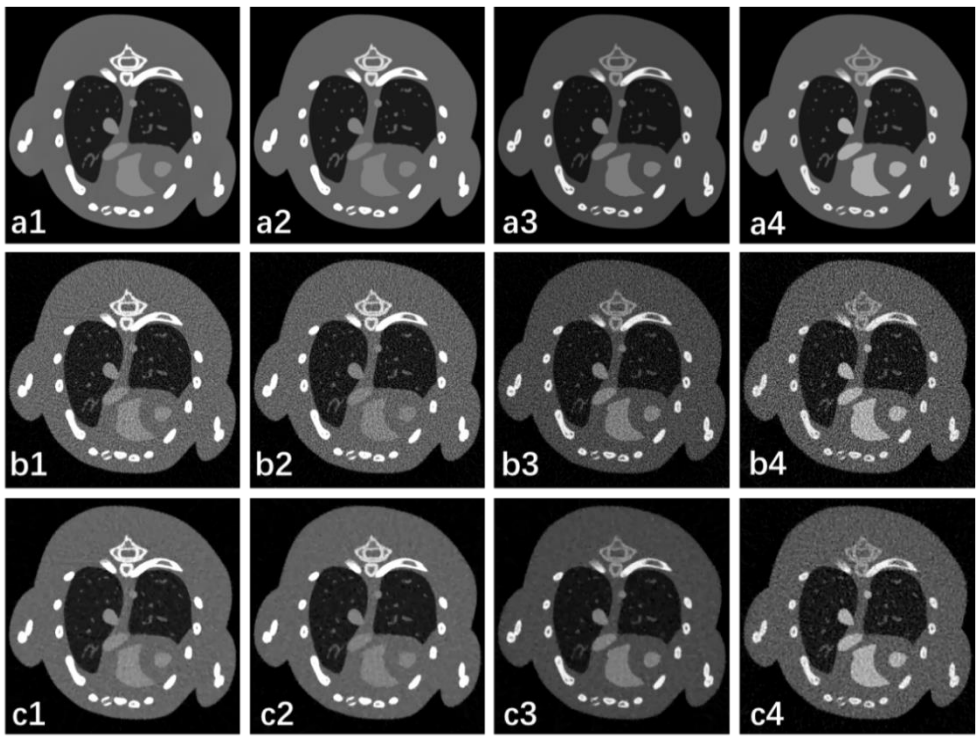

**Figure 3. Reconstructed spectral CT images for the numerical mouse thorax phantom. (a1)-(a4) and (b1)-(b4) represent the reconstructed images of 1st-4th energy bins using the SART from noise-free and noisy projections, respectively. (c1)-(c4) are reconstructed results using the TVM from noisy projections. The display windows of 1st-4th energy bins are [0 2] cm$^{-1}$, [0 1.2] cm$^{-1}$, [0 1.2] cm$^{-1}$ and [0 0.8] cm$^{-1}$, respectively.**

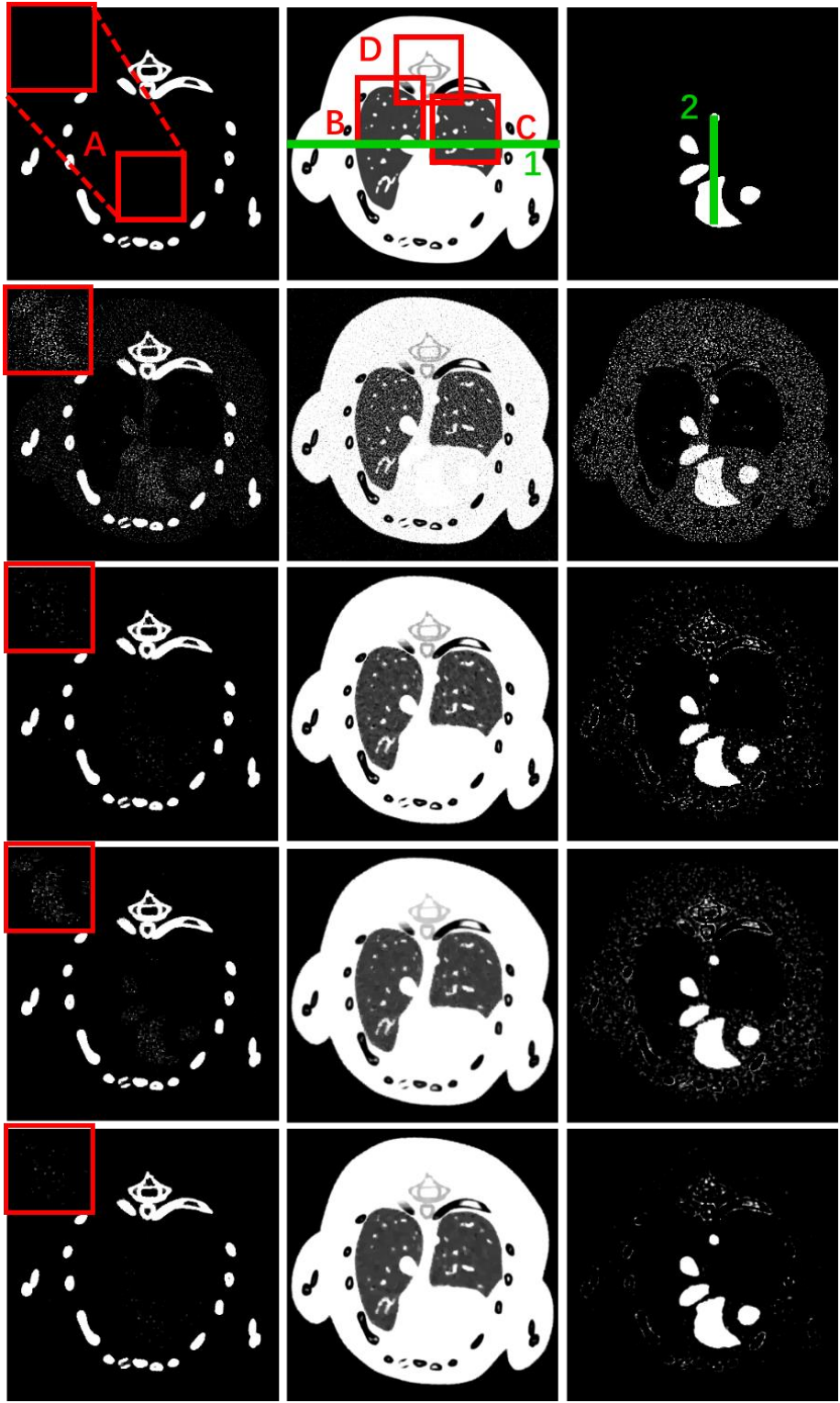

**Figure 4. Material decomposition results of the numerical mouse. The 1st-3rd columns represent bone, soft tissue and iodine contrast agent, where the display windows are [0.03 0.2], [0.10.85] and [0.0007 0.003]. The 1st row represent the ground truth, and 2nd -5th rows are material decomposition results using SART-DI, TVM-DI, SART-TVMD and TVM-TVMD methods, respectively.**

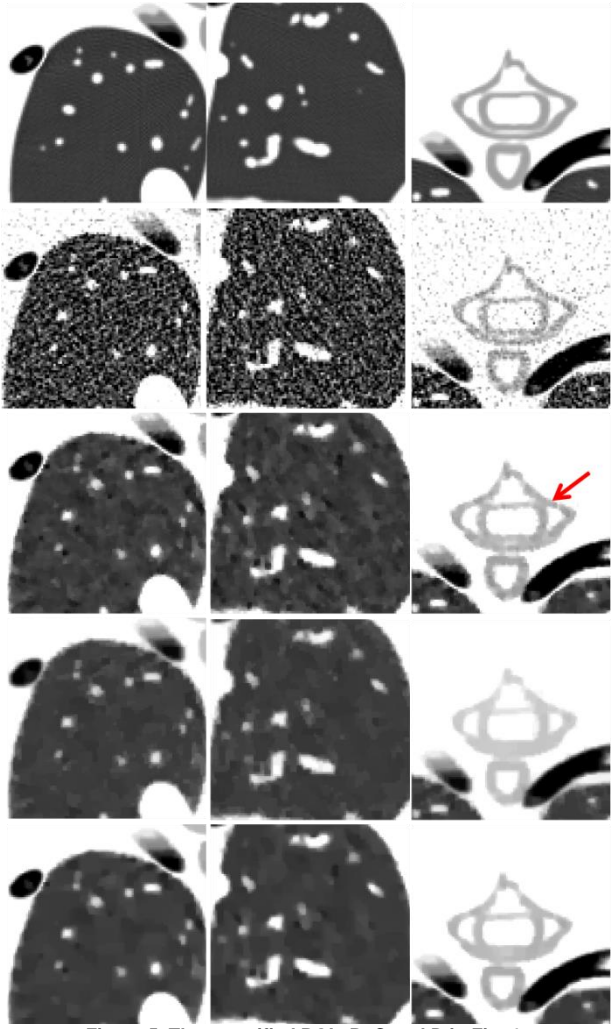

**Figure 5.** The magnified ROIs B, C and D in Fig. 4.

To compare the accuracy of decomposed materials, representative profiles of the lines indicated by "1" and "2" in **Fig. 4** are shown in **Fig. 6**. It can be seen that the profile from the TVM-TVMD is closer to the ground truth comparing with other comparisons. It can further confirm the outperformance of the TVM-TVMD method.

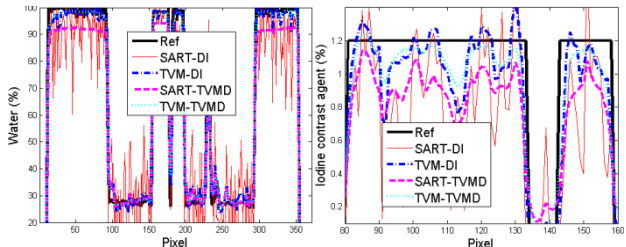

**Figure 6.** Representative profiles of line 1 (left) and 2 (right) in Fig.4.

To evaluate the performance of TVM-TVMD algorithm for improving material decomposition accuracy, the quantitative evaluation results of three basis material are given in **Table I**. From **Table I**, it can be observed that the TVM-TVMD can always achieve the smallest RMSE value with the highest PSNR and SSIM. It can further confirm that the two-step regularizations can improve the accuracy of material decomposition.

Table I. Quantitative evaluation results of three basis materials.

| | | RMSE($10^{-2}$) | PSNR | SSIM |
|---|---|---|---|---|
| Bone | SART-DI | 2.736 | 31.258 | 0.6651 |
| | TVM-DI | 1.319 | 37.5869 | 0.9573 |
| | SART-TVMD | 1.610 | 35.8651 | 0.8848 |
| | TVM-TVMD | **1.247** | **38.0799** | **0.9895** |
| Soft tissue | SART-DI | 10.069 | 19.9403 | 0.4813 |
| | TVM-DI | 3.585 | 28.9598 | 0.9015 |
| | SART-TVMD | 6.510 | 23.7282 | 0.8967 |
| | TVM-TVMD | **3.182** | **29.9552** | **0.9492** |
| Iodine contrast agent | SART-DI | 0.1407 | 57.0310 | 0.6885 |
| | TVM-DI | 0.06517 | 63.6994 | 0.8765 |
| | SART-TVMD | 0.08837 | 61.0744 | 0.7090 |
| | TVM-TVMD | **0.05639** | **64.9754** | **0.8955** |

The TVM-TVMD model contains regularization in both image reconstruction and material decomposition. In this study, we first compare the convergence of data fidelity terms with respect to iteration number for the SART and TVMD algorithms in **Fig. 7 (a)** and **(b)**. It can be seen that data fidelity terms of image reconstruction and material decomposition can converge after 15 and 6 iterations, respectively. Then, we further numerically evaluate the convergence of all methods. Particularly, **Fig. 7 (c)** and **(d)** show the RMSEs from the reconstructed images and bone component vs iteration number. This further confirms that it become stable after 15 and 6 iterations for TVM and iterative material decomposition methods, respectively.

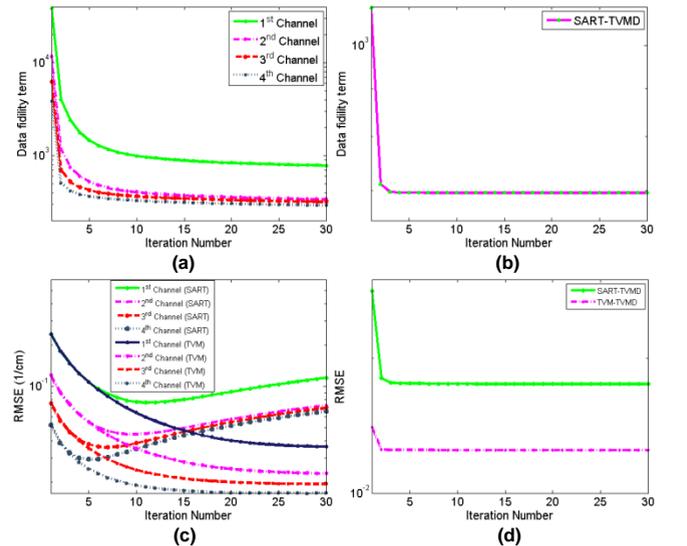

**Figure 7.** Convergent curves. (a) represents the convergent curves of data fidelity term from different energy bins by SART vs iteration number; (b) is the convergent curves of RMSE vs iteration number for SART and TVM; (c) represents the convergent curves of data fidelity term for the SART-TVMD vs iteration number; (d) represents the convergent curves of RMSE for bone vs iteration number from the SART-TVMD and TVM-TVMD

The computational costs of all methods in this study can be divided into two parts: image reconstruction and material decomposition. All methods are implemented on a PC (i5-4210U, 4.0 GB memory) with Matlab (version 2014b). Since the SART-DI contains no regularization constraints in image reconstruction and material decomposition, it has the smallest computational costs than other comparisons. There are two regularizations in TVM-TVMD model, it can consume more computational costs than other comparisons. Specifically, the SART-DI, TVMDI, SART-TVMD and TVM-TVMD consume 1.127, 2.10, 2.56 and 3.54 hours, respectively.

### B. PRECLINICAL MOUSE EXPERIMENT

A mouse injected with contrast agent (Gold nanoparticles, GNP) was scanned by a MARS micro spectral CT system (see **Fig. 8**). The spectral CT system includes one micro-focus X-ray source and one flat-panel PCD with 2 energy bins. The distances from X-ray source to the rotational center and PCD are 158 mm and 255 mm, respectively. The length of PCD is 56.32 mm with 512 pixels, leading to a FOV with a radius of 17.35 mm. Projections with 13 energy bins are collected from multiple scans with 371 views. The reconstructed image size is $512\times512\times13$, and **Fig. 8 (b)-(g)** are six representative energy bin images (1, 3, 5, 7, 9 and 13) reconstructed by the SART.

To evaluate the performance of all algorithms in preclinical applications, material decomposition results from different algorithms are shown in **Fig. 9**. For the bony components, the bony image by the SART-DI is contaminated by noise and edge of structure is also blurred. Compared with the TVM-DI and SART-TVMD results, the TVM-TVMD can obtain higher quality of bone image. For example, the image edges indicated by arrow "1" are blurred in the TVM-DI and SART-TVMD results. However, it becomes clear in the TVM-TVMD results. Besides, the bony ROI marked with "E" is extracted and magnified in **Fig. 10**. From image structure indicated by arrow "2", it can be further observed that the TVM-TVMD can protect bony structure and edge better than other comparisons. Because there exists severe noise in projections, it can reduce the reconstructed image quality and further compromise the material decomposition accuracy. Regarding the soft tissue component, the SART-DI result has much noise with the lowest image quality. The TVM-DI can provide better image quality by providing better reconstructed spectral CT image. Compared with the TVM-DI method, the SART-TVMD can decompose much smoother material image by introducing TV prior into the material decomposition model. The TVM-TVMD can explore the advantages of both TVM-DI and SART-TVMD to provide the best results. To further demonstrate the advantages of the TVM-TVMD clearly, four ROIs "E", "F", "G" and "H" are extracted and magnified in **Fig. 10**. From the tissue structure "3", it can be seen that the TVM-TVMD can retain accurate structures with higher quality. Similarly, the image structures "4" and "5" can further confirm the conclusion. As for the GNP component of the TVM-TVMD, we can see that there are less pixels of bone and the soft tissue are wrongly classified into GNP image than others. In addition, from the extracted ROI "I" in **Fig. 9**, it can be observed that there are outlier artifacts within the SART-DI and SART-TVMD results, and they can be suppressed in the TVM-DI and TVM-TVMD results.

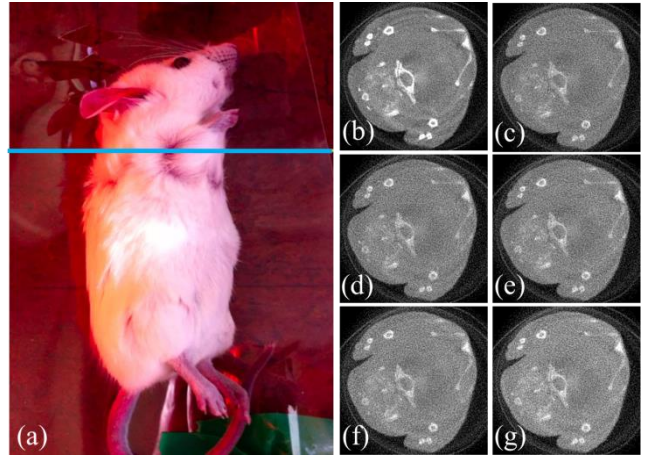

**Figure 8.** Preclinical mouse study. (a) is the mouse with injected GNP. (b)-(g) are six representative energy bin images (1, 3, 5, 7, 9 and 13) in a display window [0 0.8] cm$^{-1}$.

The parameters of two-step regularization method include the parameters in both image reconstruction and material decomposition. The parameters in the TVM mainly contain the iteration number of TV and step size of gradient descent $\alpha$. Here, the iteration number of TV is set as 20 in image reconstruction. The higher the image noise levels is, the greater the step size of gradient descent $\alpha$ value should be chosen. Because different energy bin has different noise level, different $\alpha$ should be selected respectively in practice. In this study, the x-ray energy spectrum is divided into 13 energy bins, and $\alpha$ is set as a constant 0.1 for simplification in the image reconstruction step of TVDI and TVM-TVMD. Regarding the material decomposition, the split optimization method was used in the iterative process. Regarding the parameter $\delta$, it is set as 0.001 and 0.0003 for the SART-TVMD and TVM-TVMD, respectively. Regarding the parameters in TV penalized function of the SART-TVMD method, they are set as 20, 20 and 30 for bone, soft tissue and iodine contrast agent, respectively. For the TVM-TVMD method, they are set as 200, 200 and 100, respectively. From these parameters, we can see that higher reconstructed image quality corresponds to smaller $\delta$ and greater TV penalized factor. However, we need to make a tradeoff between $\delta$ and TV penalized parameter.

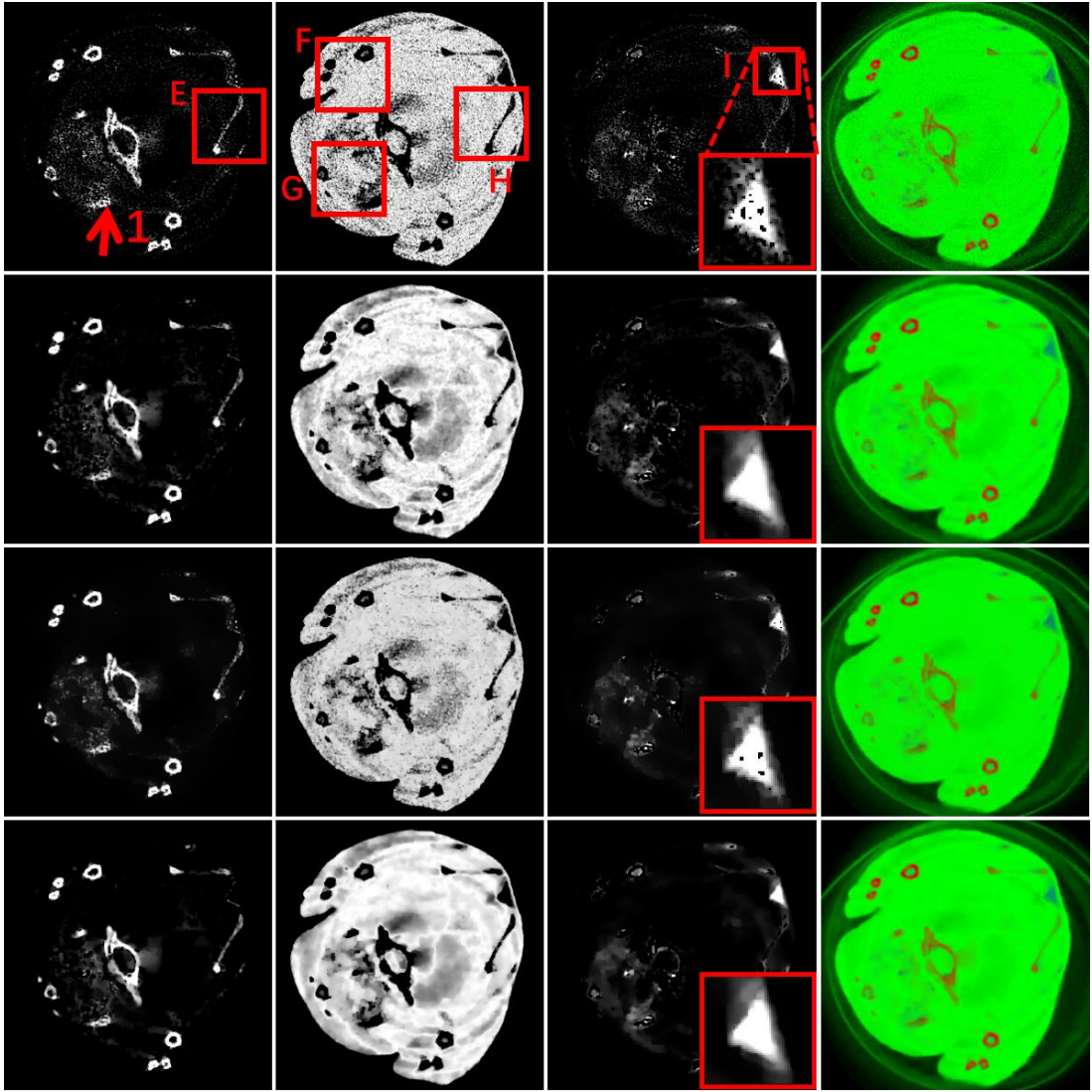

**Figure 9.** Material decomposition results from the preclinical mouse study. For top to bottom, the rows represent the SART-DI, TVM-DI, SART-TVMD and TVM-TVMD methods, respectively. The 1$^{st}$-3$^{rd}$ columns represent the bone, soft tissue and GNP in the display windows [0 0.5], [0.8 1] and [0.0 0.3], respectively. The 4$^{th}$ column images are the corresponding color rendering, where red, green and blue represent bone, soft tissue and GNP.

## IV. DISCUSSIONS AND CONCLUSIONS

To investigate the regularization priors incorporated in image reconstruction and post material decomposition models, the performance in these two cases are compared at first. To further improve the accuracy of decomposed materials, a two-step regularization strategy is developed. Particularly, a common regularization prior based on the piecewise constant assumption is chosen to implement our study. In fact, the regularization prior in image reconstruction model can obtain high quality of reconstructed images. On one hand, since the post material decomposition is a non-linear optimization problem, the improvement of material decomposition from high-quality reconstructed image is limited. On the other hand, the accuracy of decomposed materials is also limited by noise within projections as the regularization prior penalized on material decomposition results directly. The two-step regularization can inherit their advantages simultaneously to improve the accuracy of material decomposition. The numerical simulations and real mouse experiments are performed to validate the advantages of the two-step regularization based method.

In this study, we only adopt the second strategy [53] to optimize Eq. (12). To highlight the advantages of using strategy 2, the material decomposition results (including bone, soft tissue, iodine contrast iodine and air) using the SART-DI method from strategy 1 are also given in **Fig. 11**. From **Fig. 11**, it can be seen that the accuracy of material decomposition results will be degraded using strategy 1 by comparing with these results using strategy 2.

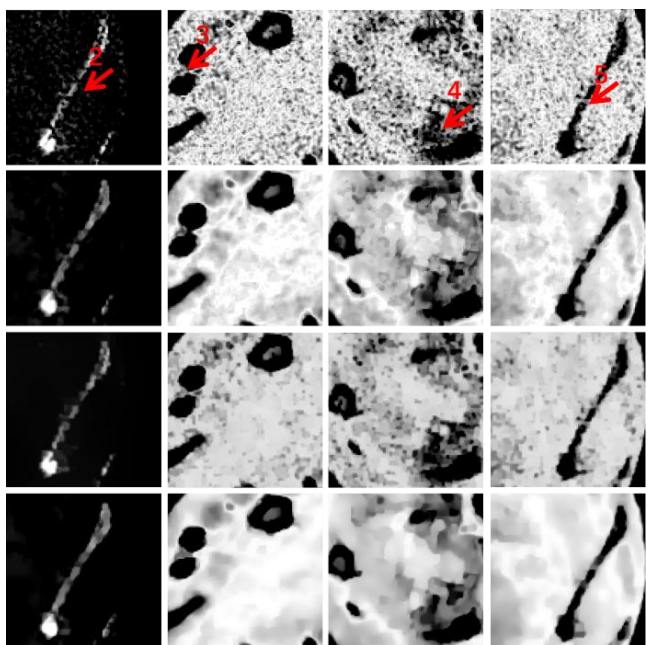

**Figure 10.** The magnified ROIs E, F, G and H in Fig. 9.

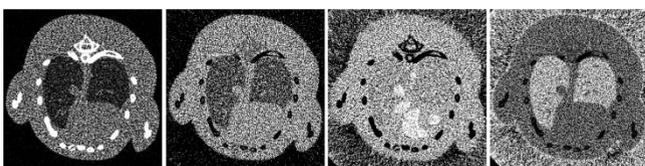

**Figure 11.** Material decomposition results of the numerical mouse by the SART-DI method. The 1st -4th columns represent bone, soft tissue, iodine contrast agent and air, where the display windows are [0.03 0.2], [0.10.85], [0.0007 0.003] and [0 1], respectively.

In this study, only the piecewise constant based prior (e.g., TV) is incorporated into the image reconstruction and material decomposition steps, and its performance is compared with single regularization based methods. Meanwhile, the tensor dictionary learning approach has been employed for spectral CT image reconstruction, and the results demonstrated tensor dictionary learning based method can achieve better performance[28] in image detail reservation and feature recovery. Similarly, it can also be employed for material decomposition. To further improve the material decomposition accuracy in real mouse experiment, the results of two-step regularization with the tensor dictionary learning are given in **Fig. 12**. According to the magnified ROIs E, F, G and H in 2nd row of **Fig. 12**, we can observe that the tensor dictionary learning approach can achieve higher quality of material images and clearer image edges than the TVM-TVMD.

Compared with the regularization prior in image reconstruction or post material decomposition, the two-step regularization based method can obtain higher accuracy of material decomposition. However, there are still some issues. First, compared with the regularization prior in image reconstruction or post material decomposition, the number of parameters in the two-step regularization model is the summation of the formers, which becomes a big challenge for the two-step regularization based method in practice. Here, these parameters are only empirically optimized by numerous experiments. Third, the employed phantoms in all experiments only contain three basis materials, and the performance of all different methods in multiple materials (greater than 3) case is still an open problem.

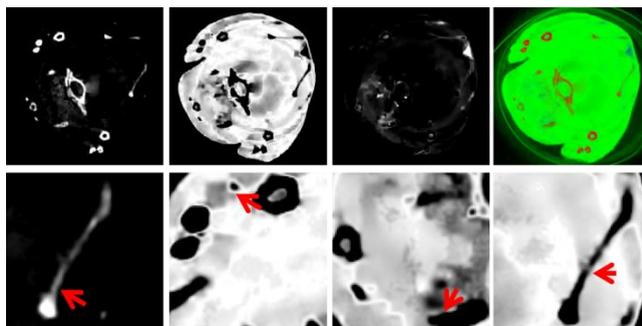

**Figure 12.** Material decomposition results from the preclinical mouse study using the two-step regularization with tensor dictionary learning in image reconstruction and material decomposition. From left to right in the 1st row, the columns represent bone, soft tissue, GNP and color rendering. From left to right in the 2nd row, the columns represent magnified ROIs E, F, G and H indicated in Fig. 9. Here, the windows for bone, soft tissue and GNP are [0 0.5], [0.8 1] and [0.0 0.3], respectively.

In conclusion, to improve the accuracy of material decomposition, a two-step regularization based method, where the TV is treated as a typical prior, was developed by comparing the performance of regularization prior in image reconstruction or post material decomposition steps. Both numerical simulations and real mouse experiments were performed to validate the advantages of the two-step regularization based method. The impact would be significant in improving accuracy of material decomposition for spectral CT.

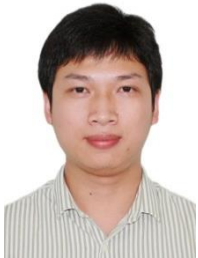

**Weiwen Wu** was born in Loudi, Hunan, China in 1991. He received the Ph.D. degree in Optoelectronic engineering from Chongqing University, Chongqing, in 2019. From 2017 to 2018, he was a Joint-training Ph.D candidate in University of Massachusetts Lowell with Dr. Hengyong Yu.

His research interests include X-ray system design, image reconstruction, dynamic bowtie and material decomposition. He is a Guest Editor of the EURASIP Journal on Advances in Signal Processing. Dr. Wu's awards and honors include the Chinese Scholarship Council, Non-destructive testing scholarship in 2019. He has published > 25 academic papers and abstracts in peer-reviewed publications.

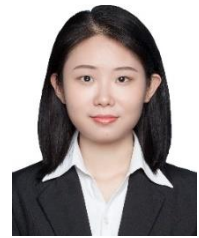

**Peijun Chen** was born in Jiangxi, China in 1994. She graduated from the College of Measurement and Optoelectronic Engineering in Nanchang Hangkong University, China, with the bachelor degree in 2016 and completed the master degree from the College of Optoelectronic Engineering in Chongqing University, China, in 2019.

Her research interests focus on CT image reconstruction, spectral CT image analysis and material decomposition.

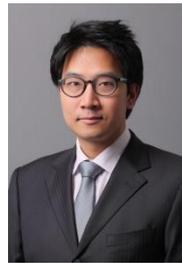

**Vince Vardhanabhuti** is Clinical Assistant Professor at the Department of Diagnostic Radiology, Li Ka Shing Faculty of Medicine, The University of Hong Kong. He completed his medical degree at Guy's, King's and St Thomas' School of Medicine at King's College, London, UK in 2005. He had subsequent training in London, Oxford, Plymouth, Exeter, and completed his Radiology training as a Fellow at Imperial College London, UK. He also completed a PhD in iterative reconstruction in computed tomography during training. He engages with various research projects relating to medical imaging, with the goal of early clinical translation to benefits patients. He serves on Scientific Editorial Board in Computed Tomography section in European Radiology. He has published > 90 academic papers and abstracts in peer-reviewed publications.

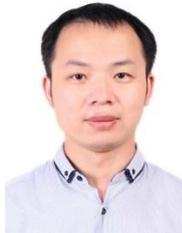

**Weifei Wu** was born in Hunan, China in 1985. He received the B.S. degrees in the medical College of Yangtze University, China, in 2010, the M.S. degree in the medical College of Nanjing University, China, in 2013 and Ph. D in the medical College of Wuhan University, China, in 2019. He is also a clinical doctor in The First People's Hospital of Yichang. In 2017, he obtained the Young Grant from National Natural Science Foundation of China.

His research interests include CT image analysis, MRI reconstruction, clinical study including mechanism and treatment of nerve injury. He is a reviewer of the Spine journal, IJS journal and so on.

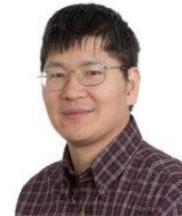

**Hengyong Yu** is Full Professor and Director of the Imaging and Informatics Lab, Department of Electrical and Computer Engineering, University of Massachusetts Lowell. He received his Bachelor's degrees in information science & technology (1998) and computational mathematics (1998) respectively, and his PhD degree in information & communication engineering (2003) from Xi'an Jiaotong University. His interests include medical imaging with an emphasis on computed tomography and medical image processing and analysis. He has authored/coauthored >150 peer-reviewed journal papers and >120 conference proceedings/abstracts. According to Google Scholar Citation, his H-index is 39 and i10-index is 96. He is the founding Editor-in-Chief of *JSM Biomedical Imaging Data Papers*, serves as an Editorial Board member for *IEEE Access*, *Signal Processing*, *CT Theory and Applications*, and so on. In January 2012, he received an NSF CAREER award for development of CS-based interior tomography.